\documentclass[final,5p,times,twocolumn]{elsarticle}

\usepackage{graphicx}
\usepackage{epsfig}
\usepackage{amssymb}
\usepackage{amsmath}
\usepackage{slashed}
\usepackage{units}
\usepackage{textcomp}
\usepackage{hyperref}
\usepackage{array}
\usepackage{lmodern}

\biboptions{sort&compress}

\journal{Physics Letters B}

\begin{document}

\begin{frontmatter}

\title{Search for passing-through-walls neutrons constrains hidden braneworlds}

\author[lps]{Micha\"{e}l Sarrazin\corref{cor}}
\ead{michael.sarrazin@unamur.be}
\author[lpsc]{Guillaume Pignol\corref{cor}}
\ead{guillaume.pignol@lpsc.in2p3.fr}
\author[lpsc]{Jacob Lamblin}
\author[lpsc]{Jonhathan Pinon}
\author[lpsc]{Olivier M\'{e}plan}
\author[larn]{Guy Terwagne}
\author[larn]{Paul-Louis Debarsy}
\author[bcrc]{Fabrice Petit}
\author[ill]{Valery V. Nesvizhevsky}

\cortext[cor]{Corresponding author}

\address[lps]{LPS-PMR, University of Namur, 61 Rue de Bruxelles, B-5000 Namur, Belgium}
\address[lpsc]{LPSC, Universit\'{e} Grenoble-Alpes, CNRS/IN2P3, 53 Avenue des Martyrs, F-38026 Grenoble, France}
\address[larn]{LARN-PMR, University of Namur, 61 Rue de Bruxelles, B-5000 Namur, Belgium}
\address[bcrc]{BCRC, 4 Avenue du Gouverneur Cornez, B-7000 Mons, Belgium}
\address[ill]{Institut Laue-Langevin, 71 Avenue des Martyrs, F-38042 Grenoble, France}

\begin{abstract}
In many theoretical frameworks our visible world is a $3$-brane, embedded in a multidimensional bulk, possibly coexisting with hidden braneworlds. Some works 
have also shown that matter swapping between braneworlds can occur. Here we report the results of an experiment -- at the Institut Laue-Langevin 
(Grenoble, France) -- designed to detect thermal neutron swapping to and from another braneworld, thus constraining the probability $p^2$ of such an event. The 
limit, $p<4.6\times 10^{-10}$ at $95 \%$ C.L., is $4$ orders of magnitude better than the previous bound based on the disappearance of stored ultracold neutrons. 
In the simplest braneworld scenario, for two parallel Planck-scale branes separated by a distance $d$, we conclude that $d>87$ in Planck length units.
\end{abstract}


\end{frontmatter}

Our observable Universe could be a braneworld: a four-dimensional surface embedded in a higher dimensional spacetime (the bulk) \cite{1,2,3,4,5,6,7}. 
This idea is frequently considered in connection with the deepest questions of fundamental physics.
In particular, quantum theories of gravity foresee the existence of extra dimensions to describe the spacetime at the Planck scale. 
Hidden braneworlds are also invoked to solve the hierarchy problem \cite{8,9,10} or to elucidate the nature of dark matter and dark 
energy \cite{11,12,13,14,15,16}.
However, empirical evidence supporting braneworlds is currently lacking. 
In this paper, we report the results of a neutron-passing-through-walls experiment designed specifically to detect neutron swapping to and from another 
braneworld \cite{19}. 
Indeed, as detailed hereafter, the theory allows for particles swapping between two adjacent branes in the bulk \cite{17,18,19}. 
We used the nuclear reactor of the Institut Laue-Langevin (Grenoble, France): a very bright source of neutrons which possibly also emits neutrons copiously into 
a hidden braneworld. To detect neutrons swapping back from the hidden world, we used a helium-3 counter shielded against the neutron background of the reactor 
hall with a rejection factor of about a million.
Even without significant excess of events in the detector, we can set a limit on the neutron - hidden neutron swapping probability. Our experiment constitutes a 
unique experimental window to braneworlds and to Planck scale physics. Beyond braneworlds, our improved bound is relevant for other new-physics scenarios 
predicting oscillations of the neutron into a sterile particle \cite{20,21,22,23,24}: quite a common concept.

\begin{figure*}
\centerline{\ \includegraphics[width=18.3 cm]{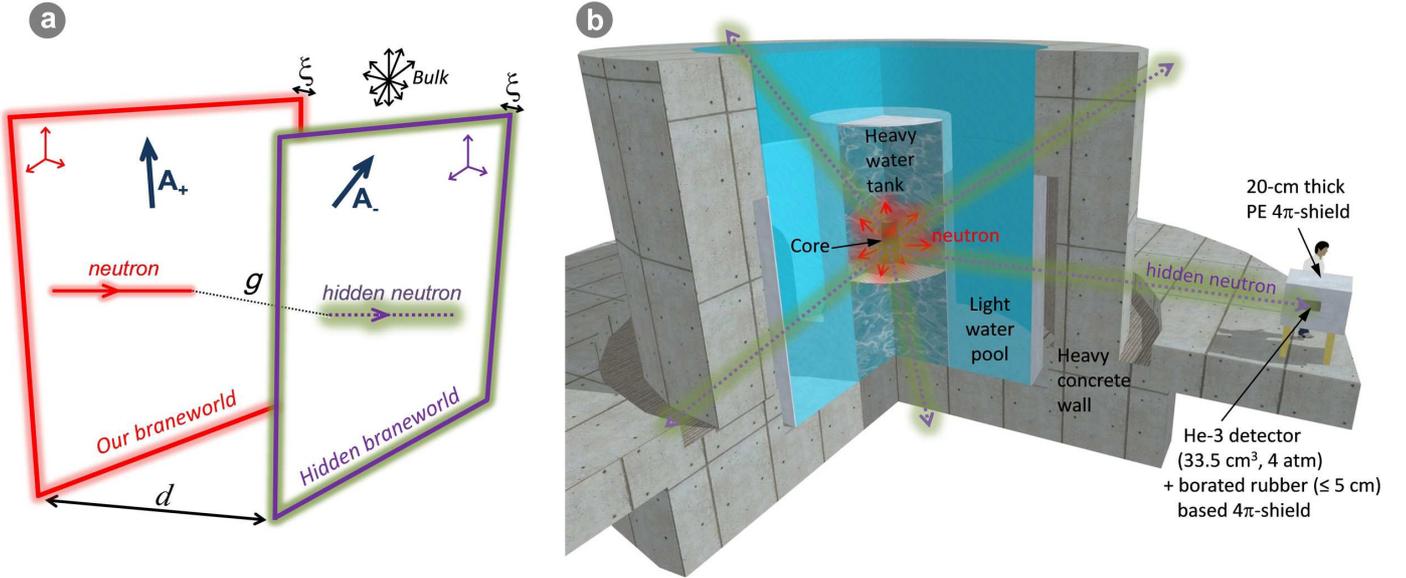}}
\caption{Scheme of the measurement. 
(a). 
Neutron swapping from our braneworld to a hidden one situated at a distance $d$ in the bulk. 
(b). 
Simplified scale diagram of the experiment at the Institute Laue-Langevin. 
The nuclear reactor (thermal power 58~MW) produces a neutron flux of about $1.5\times 10^{15}$ neutrons/s/cm$^2$. 
A compact fuel element sits in the centre of a $2.5$ m diameter tank containing the heavy water moderator. 
Details such as inserts for neutron beam lines are not shown. 
The heavy water tank serves as a source of hidden neutrons generated by the (\textit{n} + D$_2$O $\rightarrow$ hidden \textit{n} + D$_2$O) scattering 
processes \cite{19}. 
The detector is situated behind the biological shielding ($1.5$~m thick light water and $1.7$~m thick dense concrete) 
at $7.35$ m $\pm $ $0.15$ m from the centre of the reactor and at $30$ cm below the median plane. 
We used a cylindrical helium-3 ($+$ $5$ \% of CO$_2$) counter with a volume of $33.5$ cm$^3$ and a gas pressure of $4$ atm at 20$^{\circ }$C. 
The detector was surrounded by a dedicated multilayer neutron shield. 
The innermost layer is a cylindrical box made of borated rubber ($40\%$ boron content, thickness from $2$~cm to $5$~cm). 
The outermost layer is a 20-cm-thick polyethylene box.
}
\label{diag}
\end{figure*}

Let us first assume that our 3D world consists in fact of a braneworld -- a $\xi$-thick domain wall -- in a higher dimensional bulk \cite{1,6,7} (see Fig. 1a). 
Standard-Model particles are trapped along this wall which is realized in the bulk as a scalar-field soliton \cite{1}, as suggested by the effective field 
theories relating to the low-energy limit of string theory \cite{6}. 
Although many braneworlds could coexist within the bulk \cite{5,11,12,13,15,17,18,19}, in the following we consider a two-brane Universe consisting of two copies 
of the Standard Model, localized in two adjacent 3D branes (see Fig. 1a). 
While -- for processes below the brane energy scale $\hbar c / \xi$ -- these two sectors are mutually invisible to each other at the zeroth-order approximation, 
matter fields in separate branes mix at the first-order approximation through 
$\mathcal{L}_c=ig \overline{\psi }_{+}\gamma ^5\psi _{-}+ig\overline{\psi }_{-}\gamma ^5\psi_{+}$, 
where $\psi_{\pm}$ are the Dirac fermionic fields in each braneworld -- denoted $(+)$ and $(-)$ \cite{17}. 
The interbrane coupling $g$ intrinsically depends on the distance $d$ between branes and on their thicknesses $\xi$ as 
$g \sim (1/\xi)\exp(-d /\xi)$ \cite{17}. 
The brane energy scale could be as high as the Planck scale $m_{\rm Pl} \approx 10^{19}$~GeV: well beyond the reach of direct searches at high energy particle 
colliders.
Still, it is possible to explore the braneworld through matter swapping induced by the coupling $g$ at low energy. 
Indeed, precision experiments, in particular with neutrons, can be designed to monitor matter disappearance or disappearance-reappearance 
processes \cite{18,19}.

Within the nonrelativistic limit, one can show that a neutron could oscillate between two states, one localized in our brane, the other localized in the hidden 
world. 
In fact, the oscillation would be driven by the effective magnetic field $\mathbf{B}_{\bot} \mathbf{=}g\left( \mathbf{A}_{+}-\mathbf{A}_{-}\right)$ transverse to 
the branes, where $\mathbf{A}_{\pm}$ are the magnetic vector potentials in each brane (see Fig. 1a). 
Specifically, the interaction Hamiltonian $\mathbf{H}_c$ between the Pauli spinors of the visible and hidden worlds is given by \cite{15,17,18,19}:
\begin{equation}
\mathbf{H}_c=\hbar \Omega \left( 
\begin{array}{cc}
0 & \mathbf{\varepsilon } \\ 
\mathbf{\varepsilon }^{\dagger } & 0
\end{array}
\right) ,  \label{Coupling}
\end{equation}
where $\mathbf{\varepsilon }=-i\mathbf{\sigma} \cdot \mathbf{B}_{\bot}/B_{\bot}$ is a unitary matrix acting on the spin, and 
$\hbar \Omega = \mu_n B_{\bot}$, $\mu_n$ is the magnetic moment of the neutron. 
Here vector potentials $\mathbf{A}_{\pm}$ are dominated by the huge ($\sim 10^9$ T m to $10^{12}$ T m) overall astrophysical magnetic vector potential, related 
to the magnetic fields of all the astrophysical objects (planets, stars, galaxies, etc.) \cite{25,26}. 
Since the magnitude of $\left| \mathbf{A}_{+}-\mathbf{A}_{-}\right|$ is fundamentally unknown \cite{15,18}, the relevant parameter quantifying 
the coupling between the braneworlds is $B_{\bot}=g \left| \mathbf{A}_{+}-\mathbf{A}_{-}\right|$ rather than just $g$.

Due to the coupling (\ref{Coupling}) the neutron's wavefunction oscillates between the visible and the hidden states, 
at an angular frequency $\eta$ given by the energy difference between both sectors: $\eta \hbar =V_{grav,+}-V_{grav,-}$, where $V_{grav,\pm }$ are the 
gravitational potential energies felt by the neutron in each brane. 
It is likely that the energy difference is big ($\eta \gg \Omega$), resulting in very high frequency and low amplitude oscillations. 
In this case, the mean swapping probability $p$ between the visible and hidden sectors \cite{15,17,18,19} is given by: $p=2\Omega ^2/\eta ^2$.

Here, we present a neutron-passing-through-walls experiment \cite{19} (see Fig. 1b), from which we set an upper limit on the probability $p$ for a neutron to 
convert into a hidden state. 
We will then interpret the result in terms of braneworld physics.

A neutron $n$ could transform into a hidden neutron $n'$ when colliding with a nucleus. 
This process is quantified by the microscopic cross section 
$\sigma (n + {\rm nucleus} \rightarrow  n' + {\rm nucleus}) = (p/2) \sigma _s$ where $\sigma _s$ is the normal elastic scattering cross section. 
From a practical point of view, each collision at a nucleus acts as a quantum measurement and the neutron is reduced either in our, visible, world or in the other, 
invisible, braneworld (to become a hidden neutron) with a probability $p/2$ \cite{19}.
Hidden neutrons could, therefore, be generated in the moderator medium of a nuclear reactor, where a high flux of neutrons undergoes many elastic collisions. 
Being located in another braneworld, these hidden neutrons would interact very weakly with matter and freely escape the reactor. 
However, the reverse swapping process would permit us to detect them -- with an efficiency also proportional to $p$ -- 
using a usual neutron detector located close to the reactor. 
The disappearance and reappearance of neutrons due to the swapping between braneworlds would lead to the possibility of neutrons passing through a wall. 

More precisely, if we know the map of the neutron flux $\Phi _{+}(\mathbf{r})$ inside the moderator of a nuclear core we can calculate the source term 
$S_{-}(\mathbf{r})$; it corresponds to the number of hidden neutrons generated per unit volume and unit time \cite{19}: 
\begin{equation}
S_{-}(\mathbf{r}) = \frac 12 \, p \, \Sigma _s \, \Phi _+(\mathbf{r}),  \label{s}
\end{equation}
where $\Sigma _s$ is the macroscopic cross section for elastic scattering. The latter is obtained by multiplying the microscopic cross section $\sigma_s$ by the 
number density of nuclei in the moderator. 
Then, the hidden neutron flux $\Phi _{-}$ at the position $\mathbf{r_d}$ of a detector outside the reactor is given by the standard expression for an extended 
source integrated over the moderator volume: 
\begin{equation}
\Phi _{-}(\mathbf{r_d})= \frac{1}{4\pi} \int_{\rm Tank} S_{-}(\mathbf{r})/\left| \mathbf{r-r_d}\right| ^2d^3r.  
\label{phim}
\end{equation}

For our experiment, we used the High Flux Reactor of the Institut Laue-Langevin (details are in Fig. 1b). 
We got $\Phi _{+}(\mathbf{r})$ from a numerical computation using MURE 
(MCNP Utility for Reactor Evolution) \cite{27} a Monte Carlo N-Particle transport code (MCNP) \cite{28} 
coupled with fuel burnup calculations.
The core dynamics were modelled assuming a simplified geometry (i.e. uniform fuel, minor neutron beam-tubes omitted, etc.) 
From this simulation, we obtained the evolution of the thermal neutron flux distribution after the start of the reactor cycle 175 (17$^{th}$ June 2015) and 
during the experiment (6-10$^{th}$ July 2015). 
In order to estimate the accuracy of the calculation, the simplified dynamic simulation was compared with a static simulation, using a fully detailed 
geometry made by ILL staff \cite{29}. 
We found that the discrepancy in the total neutron fluxes is lower than $2\%$ but the dynamical model 
overestimates the thermal neutron flux at the periphery of the reactor by up to $20\%$.
To get a conservative estimate of the hidden neutron flux, we applied a global reduction factor of $0.8$ of the neutron flux $\Phi_{+}$ calculated with the 
dynamical model. 
Finally, the hidden neutron flux $\Phi_{-}$ at the position of the detector can be computed using equations (\ref{s}) and (\ref{phim}). 
In the calculation, we considered only the flux of thermal neutrons and neglected the small contribution of epithermal and fast neutrons. 

To regenerate and detect hidden neutrons in our visible brane, we used a helium-3 gaseous detector (see details in Fig. 1b). 
For monochromatic neutrons, the event rate $\Gamma$ detected in our brane is \cite{19}: 
\begin{equation}
\Gamma =\frac 12 \, p \, \Sigma _A \, \Phi _{-}(\mathbf{r_d}) \, V,  \label{gamma}
\end{equation}
where $V$ is the volume of the detector and $\Sigma _A$ is the macroscopic absorption cross section of helium-3. 
For a continuous energy spectrum, the event rate is obtained by integrating the equation over the spectrum. Of course, 
$\Gamma$ is directly constrained from the measurements.

Suppressing the background of the detector constitutes the major challenge of this kind of experiment 
because the detector is installed in the reactor hall where the residual neutron flux leaking from neighbouring instruments is quite high. 
We protected the detector from background neutrons with a multilayer shield. 
The innermost layer (a cylindrical box made of borated rubber) has a high efficiency for capturing thermal neutrons. 
The outermost layer (a 20-cm-thick polyethylene box) thermalizes the background of fast and epithermal neutrons. 
Without any shielding, the detector counts an activity of $60$ c/s. 
With the outermost shield alone, the count rate drops to $0.3$ c/s. 
Using the full shielding assembly, we measured $(8.5 \pm 2.4) \times 10^{-5}$ c/s. 
The constraint on hidden neutrons will be derived from this last configuration. 
Details of the analysis are given in Fig. 2.

\begin{figure}[t]
\centerline{\ \includegraphics[width=8.9 cm]{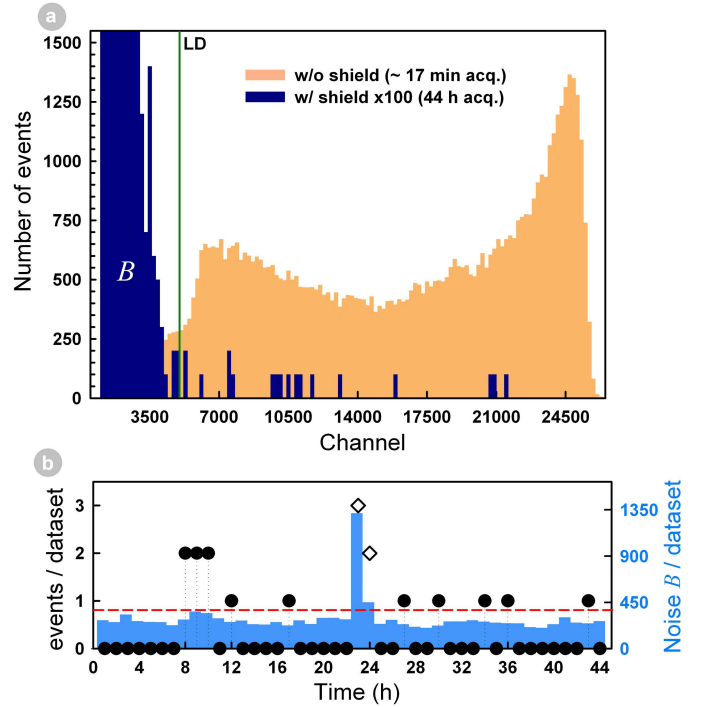}}
\caption{Experimental data. 
(a). Measured spectra with (dark blue) and without (orange) the shield. 
The spectrum shape is typical and explained by the energy deposited by the triton and proton emitted after the neutron capture by helium-3. 
LD (left green vertical line) is the lower discriminator for rejection of gamma rays and electronic noise. 
The area $B$, i.e., the full number of counts below LD, allows us to quantify the noise. 
(b). Number of events per dataset as a function of time with neutron shield (black bullets and white diamonds). 
A dataset corresponds to a 1 hour long acquisition period. 
Blue bars: noise $B$ for each dataset. 
A transient noise is observed around hour 23 -- probably due to the activity in neighbouring experiments. 
Since a large noise can mimic a neutron signal, we removed noisy datasets (marked as diamonds) using a threshold on $B$ (the horizontal red dashed line). 
}
\label{result}
\end{figure}

We interpret the recorded rate in the detector as the sum of two positive contributions: 
$\Gamma_{mes} = \Gamma_b + \Gamma$, where 
$\Gamma_b$ is the rate due to the background (internal alpha background, residual thermal or epithermal neutrons leaking through the shield and coming from the neighbouring experiments or induced by cosmic muons, electronic noise and gamma rays), 
and $\Gamma = \Gamma \left( n\rightarrow n' \rightarrow n\right)$ is the rate of hidden neutrons given by equation (\ref{gamma}). 
As a result, the data allow us to set an upper constraint on the hidden neutron $n'$ detection, using Poisson statistics:
\begin{equation}
\Gamma \left( n\rightarrow n' \rightarrow n\right) <1.37\times 10^{-4} \text{ s}^{-1}\text{ at 95\% C.L.}  \label{gammaexp}
\end{equation}

The nonzero detected rate must not be considered evidence for hidden neutrons. 
Indeed, we cannot exclude neutrons leaking through the shielding or secondary particle creation in the device. 
For instance, fast neutrons that have been insufficiently thermalized by the PE shield could have entered the detector. 
The residual counting rate was too low to identify the origin of these events precisely. To understand these events, a longer acquisition time would be 
necessary, as well as measurements with reactor off, in addition to specific simulations. This issue will be considered in further work.

\begin{figure}[t]
\centerline{\ \includegraphics[width=8.9 cm]{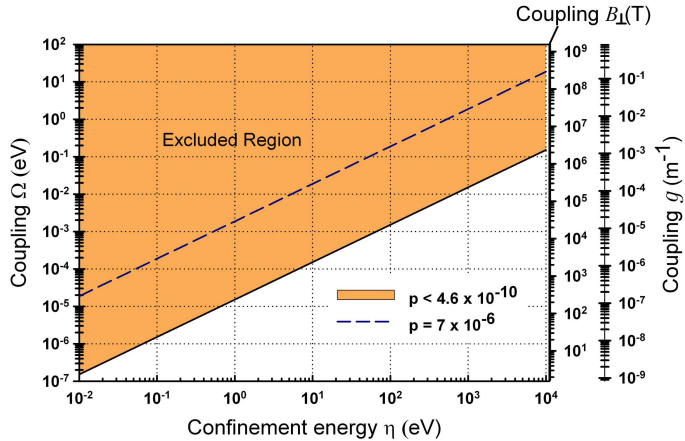}}
\caption{Regions of exclusion in the $\Omega$ - $\eta$ plane.
The orange domain is ruled out by our experiment. 
Blue dashed curve: limit from previous work \cite{18}. 
For the limit on the intrinsic interbrane coupling $g$ we assumed a conservative value of 
$\left| \mathbf{A}_{+}-\mathbf{A}_{-}\right| = 2\times 10^9$ T m \cite{25,26}.}
\label{const}
\end{figure}

From the limit given by the inequality (\ref{gammaexp}), we then extracted the upper limit on $p$ from the computed 
neutron flux $\Phi _{+}$ inside the moderator tank, and by using equations (\ref{s}) to (\ref{gamma}):
\begin{equation}
p<4.6\times 10^{-10} \text{ at 95\% C.L.}  \label{p}
\end{equation}
Our limit is better by a factor $15,000$ compared with the previous work \cite{18} (which was based on the disappearance of stored ultracold neutrons). 
Alternatively, we can derive a limit on the hidden neutron production cross section for neutron -- deuteron scattering in heavy water:
\begin{equation}
\sigma(n+D\rightarrow n'+D) <3.5 \text{ nb} \text{ at 95\% C.L.}  \label{cs}
\end{equation}

Let us now discuss the constraint given by the inequality (\ref{p}) in terms of the braneworlds' parameters. 
Fig. 3 shows the corresponding exclusion limits on the coupling $\Omega$ (or, equivalently the interbrane magnetic field $B_{\bot}$) as a 
function of the energy difference $\eta \hbar$. 
In addition, we plotted a constraint on the interbrane coupling $g$ that was deduced with conservative assumptions about the magnitude of the magnetic vector potential. 

It is instructive to interpret these limits in the most naive model of a multi-brane Universe \cite{17} with a simplifying hypothesis. 
Assuming that the hidden brane is empty, with zero gravitational energy, we set $\eta \hbar \approx 100$~eV corresponding to the typical binding energy of a neutron in 
the galactic gravitational field \cite{15}. 
Then we assume that the thickness of the brane is given by the Planck length $\xi = L_P \approx 10^{-35}$ m. 
In the simple model, the interbrane coupling depends upon the distance $d$ between branes as $g \sim (1/\xi)\exp(-d /\xi)$ \cite{17}. 
In this scenario we conclude that $d > 87 L_P$. 
For TeV-scale braneworlds ($\xi = L_{\rm TeV} \approx 10^{-19}$ m), the limit would be $d > 50 L_{\rm TeV}$. 
Therefore, if it exists, the hidden braneworld must be relatively distant from our own visible world in the bulk. 

Our null result from the search for neutrons passing through a wall can be generalized to constrain other theories predicting the existence of hidden -- or 
sterile -- neutrons. Thus, our experiment probes any scenario with the low energy phenomenology described by the generic Hamiltonian (\ref{Coupling}), independent of the 
nature of the hypothetical sterile state. 
For instance, a sterile neutron \cite{20,21,22} living in our spacetime in the context of a bigravity approach \cite{30} -- ensuring high values of $\hbar \eta $ 
-- would be subject to the phenomenology described here.
Generic constraints on exotic neutron processes are also of importance for cosmology, in particular in the context of Big Bang Nucleosynthesis \cite{23,24}. 

\section*{Acknowledgements}
M.S. and G.T. acknowledge Louis Lambotte for his technical support. The
authors are grateful to the STEREO and GRANIT collaborations for their material and technical support. This work 
is supported by the Department of Physics of the University of Namur and the Belgian 
Federal Science Policy Office (BELSPO) through the Trans-National Neutron Initiative
(TRANSNI).\\


\end{document}